\documentclass[
    ,final            
  ]
  {aipproc}

\layoutstyle{8x11double}

\usepackage{amsmath,amssymb}
\usepackage{graphicx,epsfig,natbib}

\def \beq {\begin{equation}}
\def \edq {\end{equation}}
\def \bes {\begin{subequations}}
\def \eds {\end{subequations}}
\def \wtG {\widetilde{G}}
\def \wtS {\widetilde{S}}
\def \wtC {\widetilde{C}}
\def \calo {{\cal{O}}}
\def \bs {\bar{s}}
\def \up {\uparrow}
\def \down {\downarrow}

\begin{document}

\title{Spin-current noise from fluctuation relations}


\author{Jong Soo Lim}{address={Institut de F\'{\i}sica Interdisciplin\`aria i Sistemes Complexos IFISC (UIB-CSIC), E-07122 Palma de Mallorca, Spain}}

\author{David S\'anchez}{address={Institut de F\'{\i}sica Interdisciplin\`aria i Sistemes Complexos IFISC (UIB-CSIC), E-07122 Palma de Mallorca, Spain},
altaddress={Departement de F\'{\i}sica, Universitat de les Illes Balears, E-07122 Palma de Mallorca, Spain}}

\author{Rosa L\'opez}{address={Institut de F\'{\i}sica Interdisciplin\`aria i Sistemes Complexos IFISC (UIB-CSIC), E-07122 Palma de Mallorca, Spain},
altaddress={Departement de F\'{\i}sica, Universitat de les Illes Balears, E-07122 Palma de Mallorca, Spain}}

\begin{abstract}
We present fluctuation relations that connect spin-polarized current and noise in mesoscopic conductors.
In linear response, these relations are equivalent to the fluctuation-dissipation theorem that relates
equilibrium current--current correlations to the linear conductance. More interestingly, in the weakly
nonlinear regime of transport, these relations establish a connection between the leading-order
rectification spin conductance, the spin noise susceptibility and the third cumulant of spin current fluctuations at
equilibrium. Our results are valid even for systems in the presence of magnetic fields and coupled to ferromagnetic electrodes. 
\end{abstract}

\maketitle

We have recently derived fluctuation relations valid beyond the linear response regime of mesoscopic transport
fully taking into account the spin degree of freedom and magnetic interactions \cite{lop12}.
These relations represent higher-order fluctuation-dissipation relations \cite{esp09} and are satisfied
even if microreversibility is broken due to an externally applied magnetic field $B$ \cite{for08,san09}.

Consider a generic multiterminal conductor coupled to reservoirs $\alpha,\beta\ldots$ with voltages
$V_\alpha,V_\beta\ldots$ and background temperature $T$ (see Fig. 1). We assume that there exists a spin quantization axis common
to all terminals and thus denote the spin index with $s,s'\ldots$, where $s=+$ ($s=-$)
for electrons with spin $\uparrow$ ($\downarrow$). Additionally, the leads can be magnetized
with a polarization $p$, allowing for spin-dependent transport across the sample.
We allow for possible spin biases present in the lead voltages: $V_{\alpha s}, V_{\beta s'}\ldots$
Then, the current $\langle I_{\alpha s}\rangle$ flowing through terminal $\alpha$ associated with electronic spins $s$
can be expanded in powers of voltage around the equilibrium point:
\bes\label{eqi}
\begin{align}
\langle I_{\alpha s}\rangle &= \sum_{\beta s'} G_{\alpha s,\beta s'}^{(1)} V_{\beta s'} + \frac{1}{2}\sum_{\beta s',\gamma s''} G_{\alpha s,\beta s'\gamma s''}^{(2)}V_{\beta s'}V_{\gamma s''}\nonumber \\ &+ \calo(V^3)\,,\\
\langle S_{\alpha s\beta s'}\rangle &= S_{\alpha s\beta s'}^{(0)} + \sum_{\gamma s''} S_{\alpha s\beta s',\gamma s''}^{(1)} V_{\gamma s''} + \calo(V^2)\,, \label{eqs}
\\
\langle C_{\alpha s\beta s'\gamma s''}\rangle &= C_{\alpha s\beta s'\gamma s''}^{(0)} + \calo(V)\,, \label{eqc}
\end{align}
\eds
where $G_{\alpha s,\beta s'}^{(1)}$ is a linear conductance coefficient and
$G_{\alpha s,\beta s'\gamma s''}^{(2)}$ is the leading-order rectification term.
Both coefficients are evaluated at equilibrium but their functional dependence on
the scattering properties of the conductor is very different \cite{chr96}.
In Eqs.\ \eqref{eqs} and \eqref{eqc} we have also expanded the noise $S$ (current--current correlations)
and the third cumulant $C$, for which $(0)$ label equilibrium responses.

\begin{figure}
  \includegraphics[width=.45\textwidth]{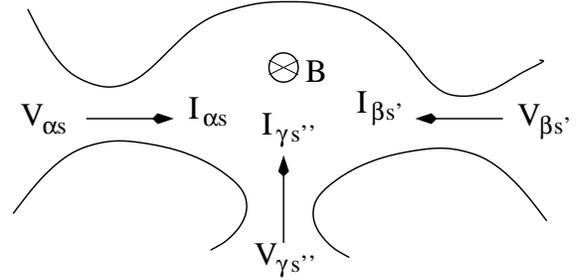}
 \caption{Sketch of a mesoscopic conductor in the presence of magnetic field $B$
 and attached to multiple terminals with applied voltages $V_{\alpha s}, V_{\beta s'}\ldots$.
 We indicate spin currents flowing through the system.}
\end{figure}

Let $f_{\alpha s\pm}=f_{\alpha s}(B,p)\pm f_{\alpha \bs}(-B,-p)$ be the symmetric ($+$) and antisymmetric ($-$) combinations
of a given transport coefficient ($G$, $S$ or $C$). In Ref.\ \cite{lop12} we show that the fluctuation relations
for spintronic systems read 
\bes\label{eqs0c0}
\begin{align}
S_{\alpha s\beta s'\pm}^{(0)} &= k_BT\left( G_{\alpha s,\beta s'\pm}^{(1)} + G_{\beta s',\alpha s\pm}^{(1)}\right)\,, \label{eqs0}\\
C_{\alpha s\beta s'\gamma s'' \pm}^{(0)} &= k_BT \left[ \left(S_{\alpha s\beta s',\gamma s''\pm}^{(1)} + S_{\alpha s\gamma s'',\beta s'\pm}^{(1)} + S_{\beta s'\gamma s'',\alpha s\pm}^{(1)}\right) \right. \nonumber \\
&\left. -k_BT\left(G_{\alpha s,\beta s'\gamma s''\pm}^{(2)} + G_{\beta s',\alpha s\gamma s''\pm}^{(2)} + G_{\gamma s'',\alpha s\beta s'\pm}^{(2)}\right) \right]\,.
\end{align}
\eds
These expressions are to be supplemented with $S_{\alpha s\beta s'}^{(0)}(B,p) = S_{\alpha\bs\beta\bs'}^{(0)}(-B,-p)$
and $C_{\alpha s\beta s'\gamma s''}^{(0)}(B,p) = -C_{\alpha\bs\beta\bs'\gamma\bs''}^{(0)}(-B,-p)$,
which state that the even (odd) equilibrium cumulants respond symmetrically (antisymmetrically)
to simultaneous reversals of the spin direction, the magnetic field and the lead magnetization. Using the symmetry
of $S_{\alpha s\beta s'}^{(0)}$ in Eq.\ \eqref{eqs0} we recover the Onsager-Casimir reciprocity relations,
valid in the linear response regime, i.e., $G_{\alpha s,\beta s'}^{(1)} (B,p)=G_{\beta \bs',\alpha \bs}^{(1)} (-B,-p)$.
Importantly, Eq.\ \eqref{eqs0c0} is valid not only for arbitrary Coulomb interactions but also
for magnetic couplings such as spin-flip relaxation processes \cite{lop12}.

Now, in spintronics it is convenient to work with {\em spin-polarized} current and noises.
In fact, spin-resolved shot noise can provide valuable information about the role of interactions
in mesoscopic systems \cite{sau04}. Therefore, we define the spin-polarized ($SP$) quantities
\bes\label{eqisp}
\begin{align}
\langle I_{\alpha}^{SP} \rangle &= \langle \left(I_{\alpha\up} - I_{\alpha\down}\right)\rangle = \langle I_{\alpha\up}\rangle - \langle I_{\alpha\down}\rangle
= \sum_s s \langle I_{\alpha s}\rangle\,, \\
\langle S_{\alpha\beta}^{SP}\rangle &= \langle \left(I_{\alpha\up}-I_{\alpha\down}\right)\left(I_{\beta\up}-I_{\beta\down}\right)\rangle
= \sum_{s,s'} ss' \langle S_{\alpha s\beta s'}\rangle \,, \\
\langle C_{\alpha\beta\gamma}^{SP}\rangle 
&= \langle\left(I_{\alpha\up}-I_{\alpha\down}\right)\left(I_{\beta\up}-I_{\beta\down}\right)\left(I_{\gamma\up}-I_{\gamma\down}\right)\rangle\nonumber\\
&= \sum_{s,s',s''} ss's''\langle C_{\alpha s\beta s'\gamma s''}\rangle\,,
\end{align}
\eds
and, similarly to Eq. \ \eqref{eqi}, expand Eq. \eqref{eqisp} in powers of voltage:
\bes
\begin{align}
\langle I_{\alpha}^{SP}\rangle &= \sum_{\beta s'} \wtG_{\alpha,\beta s'}^{(1)} V_{\beta s'} + \frac{1}{2}\sum_{\beta s',\gamma s''} \wtG_{\alpha,\beta s'\gamma s''}^{(2)}V_{\beta s'}V_{\gamma s''} \nonumber \\
&+ \calo(V^3)\,,
\\
\langle S_{\alpha\beta}^{SP}\rangle &= \wtS_{\alpha\beta}^{(0)} + \sum_{\gamma s''} \wtS_{\alpha\beta,\gamma s''}^{(1)} V_{\gamma s''} + \calo(V^2) \,,
\\
\langle C_{\alpha\beta\gamma}^{SP}\rangle &= \wtC_{\alpha\beta\gamma}^{(0)} + \calo(V)\,.
\end{align}
\eds

Our goal is to find the fluctuation relations obeyed by the response coefficients $\wtG$, $\wtS$ and $\wtC$.
We substitute Eq. \eqref{eqi} in Eq. \eqref{eqisp} and use the fluctuation relations expressed in Eq. \eqref{eqs0c0}.
Then, we find
\bes\label{eqfrsp}
\begin{align}
\wtS_{\alpha\beta\pm}^{(0)} &= k_BT \sum_s s\left( \wtG_{\alpha,\beta s\pm}^{(1)} + \wtG_{\beta,\alpha s\pm}^{(1)}\right) \label{eqwts}\\
\wtC_{\alpha\beta\gamma\pm}^{(0)} &= k_BT \sum_s s\left(\wtS_{\alpha\beta,\gamma s\pm}^{(1)} + \wtS_{\alpha\gamma,\beta s\pm}^{(1)} + \wtS_{\beta\gamma,\alpha s\pm}^{(1)}\right)\nonumber \\
&-(k_BT)^2 \sum_{s,s'} ss'\left(\wtG_{\alpha,\beta s\gamma s'\pm}^{(2)} + \wtG_{\beta,\alpha s\gamma s'\pm}^{(2)} + \wtG_{\gamma,\alpha s\beta s'\pm}^{(2)}\right)\label{eqwtc}
\end{align}
\eds
together with $\wtS_{\alpha\beta -}^{(0)}=0$ and $\wtC_{\alpha\beta\gamma +}^{(0)}=0$
(the former is consistent with the Onsager-Casimir relations). Two remarks are in order.
First, Eq.\ \eqref{eqwts} establishes a relation between the spin-resolved noise at equilibrium (i.e., in the absence of bias voltage)
and the polarized linear conductance. It represents an extension of the fluctuation-dissipation theorem
valid for spintronic systems in the presence of magnetic fields and ferromagnetic electrodes.
Second, Eq.\ \eqref{eqwtc} describes a nontrivial connection between the third cumulant at equilibrium,
the spin-resolved noise susceptibility [$S^{(1)}$] and the polarized nonlinear conductance to leading order.
Clearly, the differential conductance need not obey the Onsager-Casimir symmetry relations \cite{san04,spi04}.
This is due to breakings of microreversibility away from equilibrium. However,
Eq.\ \eqref{eqwtc} connect the different asymmetries of the noise susceptibility and the nonlinear
conductance even in systems for which $\wtC_{\alpha\beta\gamma +}^{(0)}=0$.

In conclusion, we have formulated novel fluctuation relations applied to spin-dependent current and noises
for multiterminal systems in the presence of Coulomb interactions, magnetic fields and ferromagnetic electrodes.
Our formalism is based on rather broad assumptions: equilibrium microreversibility, global detailed balance and
probability conservation \cite{lop12}. We emphasize that our resuls hold even if a spin bias is present in the system
due to possible spin accumulations formed at the boundary between the leads and the scattering region.
Therefore, our central result [Eq.\ \eqref{eqfrsp}] is an excellent tool to investigate purely spintronic
effects in out-of-equilibrium situations.


\begin{theacknowledgments}
  Work supported by the MINECO Grants No. FIS2011-23526
  and CSD2007-00042 (CPAN).
\end{theacknowledgments}



\bibliographystyle{aipproc}   


\begin{thebibliography}{99}

\bibitem{lop12}
R. L\'opez, J.S. Lim, and D. S\'anchez, \emph{Phys. Rev. Lett.} \textbf{108},
  246603 (2012).
  
\bibitem{esp09}
M. Esposito, U. Harbola, and S. Mukamel, \emph{Rev. Mod. Phys.} \textbf{81},
1665 (2009).

\bibitem{for08}
H.~F\"orster and M. B\"uttiker, \emph{Phys. Rev. Lett.} \textbf{101}, 136805 (2008).

\bibitem{san09}
D. S\'anchez, \emph{Phys. Rev. B} \textbf{79}, 045305 (2009).

\bibitem{chr96}
  T. Christen and M. B\"uttiker, \emph{EPL} \textbf{35}, 523 (1996).
  
\bibitem{sau04}
O. Sauret and D. Feinberg, \emph{Phys. Rev. Lett.} \textbf{92},
  106601 (2004).

\bibitem{san04}
D. S\'anchez and M. B\"uttiker, \emph{Phys. Rev. Lett.} \textbf{93},
  106802 (2004); \emph{Phys. Rev. B} \textbf{72}, 201308(R) (2005).
  
  \bibitem{spi04}
B. Spivak and A. Zyuzin, \emph{Phys. Rev. Lett.} \textbf{93},
  226801 (2004).

\end{thebibliography}



\end{document}